\documentclass[twoside]{article}
\usepackage{qic}
\usepackage{enumerate}

\begin{document}
\setlength{\textheight}{8.0truein}    

\runninghead{Title  $\ldots$}
            {Author(s) $\ldots$}

\normalsize\textlineskip \thispagestyle{empty}
\setcounter{page}{1}

\copyrightheading{0}{0}{2003}{000--000}

\vspace*{0.88truein}

\alphfootnote

\fpage{1}

\centerline{\bf
ON THE SECURITY OF $\alpha\eta$: RESPONSE TO}
\vspace*{0.035truein} \centerline{\bf `SOME ATTACKS ON
QUANTUM-BASED CRYPTOGRAPHIC PROTOCOLS'} \vspace*{0.37truein}
\centerline{\footnotesize
Horace P. Yuen\footnote{yuen@eecs.northwestern.edu},
 Ranjith Nair, Eric Corndorf, Gregory S. Kanter, and
Prem Kumar} \vspace*{0.015truein} \centerline{\footnotesize\it
Center for Photonic Communication \& Computing,}
\baselineskip=10pt \vspace*{0.015truein}
\centerline{\footnotesize\it Department of Electrical Engineering
\& Computer Science, Department of Physics \& Astronomy,}
\baselineskip=10pt \centerline{\footnotesize\it Northwestern
University, Evanston, IL, 60208, USA.} \vspace*{0.225truein}
\publisher{(received date)}{(revised date)}

\vspace*{0.21truein}

\abstracts{ Lo and Ko in \cite{loko} have developed some attacks
on the cryptosystem called $\alpha \eta$ \cite{prl}, claiming that
these attacks undermine the security of $\alpha\eta$ for both
direct encryption and key generation.  In this paper, we show that
their arguments fail in many different ways. In particular, the
first attack in [1] requires channel loss or length of
known-plaintext that is exponential in the key length and is
unrealistic even for moderate key lengths. The second attack is a
Grover search attack based on `asymptotic orthogonality' and was
not analyzed quantitatively in [1]. We explain why it is not
logically possible to ``pull back'' an argument valid only at
$n=\infty$ into a limit statement, let alone one valid for a
finite number of transmissions $n$. We illustrate this by a
`proof' using a similar asymptotic orthogonality argument that
coherent-state BB84 is insecure for \textit{any} value of loss.
Even if a limit statement is true, this attack is \textit{a
priori} irrelevant as it requires an indefinitely large amount of
known-plaintext, resources and processing. We also explain why the
attacks in [1] on $\alpha\eta$ as a key-generation system are
based on misinterpretations of [2]. Some misunderstandings in [1]
regarding certain issues in cryptography and optical
communications are also pointed out. Short of providing a security
proof for $\alpha\eta$, we provide a description of relevant
results in standard cryptography and in the design of $\alpha\eta$
to put the above issues in the proper framework and to elucidate
some security features of this new approach to quantum
cryptography.
 }{}{}

\vspace*{10pt}

\communicate{to be filled by the Editorial}

\vspace*{1pt}\textlineskip    

\section{Introduction}
In [1], Lo and Ko describe, without quantitative calculations,
some attacks on the direct encryption protocol of [2], interpreted
by them also as a key generation scheme.  They draw the firm
conclusion that our protocol is fundamentally insecure, that these
attacks were neglected by us as they are ``outside the original
design,'' and that they ``can, to some extent, be implemented with
current technology.''  We contend that the strength and weakness
of our scheme have been totally misrepresented in [1], which does
not analyze the relevant cryptographic problems in a meaningful
framework. Although we have already commented briefly on the
attacks of \cite{loko} in \cite{yuen04} and \cite{pla05}, and some
related comments are given by Hirota et al in \cite{hirota05},
\cite{loko} is still often quoted without also referring to our
partial rejoinder. Thus, we feel it appropriate that a specific
response to [1] be made in a complete paper.  In particular, we
would like to clear up at the same time many issues in the
practical use of quantum cryptography and in the properties of
$\alpha\eta$ that have so far not been elucidated in the
literature. We do not attempt to give a complete security proof of
$\alpha\eta$ in this paper. Such a proof is not available and is
the subject of ongoing research. See \cite{nair06} for recent
results. Nevertheless, it is possible to refute the arguments of
Lo and Ko taken by themselves, and this will be the main aim of
this paper.

First of all, we note that the attacks in [1] do not contradict
our claim in [2] that $\alpha\eta$ encryption provides
\textit{exponential complexity-based} security against
known-plaintext attacks using a particular `assisted' brute-force
search. See [2] or alternatively, [4] for a more detailed
description. Although we mention the possibility of key generation
with $\alpha\eta$, we do not present an explicit scheme to do so
in [2]. The authors of [1] assume that the protocol of [2] works
without any additions or modifications for key generation, which
was not claimed by us at all. While they arrive at attacks that
purport to show that $\alpha\eta$ is insecure in the
information-theoretic sense against known-plaintext attacks
--- already believed by us to be quite possible \cite{yuen04} --- we claim that
the two attacks in [1] do not conclusively prove insecurity of any
finite-$n$ system. Proof is important in this quantum situation
because $\alpha\eta$ falls outside the class of classical
nonrandom ciphers for which known-plaintext attacks can be proved
to succeed. But perhaps more significantly, the Lo-Ko attacks are
unrealistic in the \textit{fundamental} sense of having
exponential complexity and requiring an exponential amount of
resources. In Section 2.2, we bring out the \emph{important point}
that, in contrast to other kinds of complexity, exponential
complexity offers realistic security as good as unconditional
security.

We shall explain fully our criticisms of [1] in the course of this
paper. In this introductory section, we will lay out three major
general defects in [1] which in our opinion are also implicit in
various papers on theoretical quantum cryptography. We will later
have occasion to indicate specific points where these defects
arise when we reply in detail in Section 4 to the attacks in
\cite{loko}.

In the \textit{first} place, vague qualitative arguments are often
offered as rigorous proofs, while at the same time not giving
precise conditions under which a result is claimed to be valid. In
\cite{loko}, there are even several claims made without any
argument at all. Rigorous proofs are important in quantum
cryptography because the main superiority it claims over standard
cryptography is the possibility of rigorous proof of security,
unconditional or otherwise. A more subtle point is that many
arguments, including one in [1], rely on statements valid
\textit{at} $n=\infty$ which \textit{cannot} be cast into
\textit{limiting} statements on the relevant quantities. Indeed,
limit and continuity questions at $n=\infty$ are especially subtle
in quantum mechanics owing to the nonseparable Hilbert space,
i.e., a Hilbert space with an uncountable basis, that arises when
$n=\infty$. One pitfall of such a leap of faith is illustrated in
Section 5.

\emph{Secondly}, strong claims are made with no actual numbers or
numerical ranges indicated for the validity of the results. Thus,
results are often claimed to be valid asymptotically as the number
of bits $n$ in a sequence goes to infinity, without any estimate
on the convergence rate. Such limiting results alone are of no use
to an experimentalist or designer of a \emph{real} system. As
security proofs, they offer no quantitative guarantee of any kind
on an actual realistic system where $n$ is often not even very
large. As attacks, they imply nothing about the level of
insecurity of any finite $n$ system without
convergence-rate estimates. 
Thus, showing a scheme to be insecure simply as a limiting
statement when $n \rightarrow \infty$ has no practical
implication. (See Section 4 for a complete discussion.) A related
point is with regard to the realistic significance of quantities
that vary exponentially with respect to some system parameter.
Thus, consideration of attacks, as is the case in one attack in
\cite{loko}, that succeed only when the channel-transmittance (the
output-to-input power ratio) $\eta\sim2^{-|K|}$, where $|K|$ is
the key length, is seen to be practically irrelevant by plugging
in typical numbers for $|K|$. More significantly, attacks that
require exponential resources or processing like those in
\cite{loko} are irrelevant in a fundamental sense, because the
situation \textit{cannot} be changed by technological advances,
similar to the case of unconditional security.

These points are important because security in cryptography is a quantitative
issue. For example, in quantum key generation, the exact amount of Eve's
uncertainty determines how much key is generated. To ensure that one generates
a sufficiently large key, it is not sufficient to use qualitative arguments
that are valid only at extreme limits, since they may break down quantitatively
in realistic systems.


\emph{Thirdly}, the general approach to quantum cryptography
underlying $\alpha\eta$, called `Keyed Communication in Quantum
Noise' (KCQ) \cite{yuen04}, is not well understood. In particular,
the various and distinct issues in connection with direct
encryption and key generation with (or even without) a secret key,
which have to be clearly delineated for a proper analysis, are
lumped together in \cite{loko}, generating considerable confusion
even in the context of classical cryptography.  Since our approach
is novel, this current situation is perhaps understandable. While
the full story of this field of research is still to be
understood, some clarifications can be made to clear up the
various confusions.

In addition to the above, some specific details of implementation
of $\alpha\eta$ are also misconstrued in [1]. Along with
responding to the Lo-Ko arguments, one main purpose of this paper
is to provide the proper framework for security analysis of
$\alpha\eta$, for direct encryption as well as key generation. It
is \emph{not} the purpose of this paper to provide any detailed
security analysis of $\alpha\eta$, which is a huge undertaking and
an on-going effort. However, we will indicate the many features
that make $\alpha\eta$ uniquely interesting and useful at various
places in the paper.

The plan of this paper is as follows: In Section 2, we provide an
outline of relevant results and facts in symmetric-key
cryptography, which are not well-known. Our statements on direct
encryption cryptography in this paper refer only to the
symmetric-key case, and not to public-key cryptography. In fact,
public-key cryptography is not used for encryption of data
sequences of more than a few hundred bits owing to its slow speed.
We discuss in a subsection the current knowledge regarding
security against known-plaintext attacks in standard cryptography
and discuss the concepts of a \textit{random cipher} and a
\emph{nondegenerate} cipher. Much of this subsection as well as
Appendix A are our own contributions. They contain subtle
distinctions needed to precisely state important results, and may
be regarded as providing the basic framework in which to view
known-plaintext attacks on $\alpha\eta$ or any other randomized
encryption system. In Section 3, we review our $\alpha\eta$ scheme
and the different security issues associated with its use in
direct encryption and key generation. In Section 4, the Lo-Ko
attacks and their specific criticisms are explained and responded
to, both specifically and generally in view of the above-mentioned
defects. It will be shown that their arguments are deficient in
many different ways. To illustrate the fallacy of the `asymptotic
orthogonality' argument, a `proof' that coherent-state BB84 using
a classical error-correction code is insecure for any loss, no
matter how small, is presented in Section 5. Various other
misconceptions in [1] are listed in Section 6. A brief summary of
our conclusions is given in Section 7.

\section{Cryptography}

\subsection{Direct Encryption}

We assume that the basics of symmetric-key data encryption are
known to the reader (See, e.g., \cite{stinson,massey88}). Thus,
the $n$-symbol long plaintext is denoted by the random variable
$X_n$, the corresponding ciphertext is denoted $Y_n$ and the
secret key is denoted $K$. In standard cryptography, one usually
deals with \textit{nonrandom ciphers}, namely those cryptosystems
for which the conditional entropy
\begin{equation} \label{nonrandom}
H(Y_n|K X_n)=0. \end{equation}  Thus, the plaintext and key
uniquely determine the ciphertext. In such a case, $X_n$ and $Y_n$
are usually taken to be from the same alphabet. Note that in this
paper, equations involving $n$ as a parameter are assumed to be
valid for all $n$ unless stated otherwise. Ciphers for which
Eq.(\ref{nonrandom}) is relaxed so that the same plaintext may be
mapped for a given key to many different ciphertexts, perhaps
drawn from a different alphabet than $X_n$, will be called random
ciphers. Thus, a \textit{random cipher} is defined by
\begin{equation} \label{random}
H(Y_n|KX_n) \neq 0. \end{equation} Such ciphers are called
`privately randomized ciphers' in Ref. \cite{massey88} as the
different ciphertexts $Y_n$ for a given $X_n$ are obtained by
privately (i.e., in an unkeyed fashion known only to the sender
Alice) randomizing on a specific $Y_n$. We will just call such a
cipher a random cipher (Note that `random cipher' is used in a
completely different sense by Shannon \cite{shannon49}). For both
random and nonrandom ciphers, we enforce the condition that the
plaintext be recoverable from the ciphertext and the key, i.e.,
\begin{equation} \label{decryption} H(X_n|KY_n)=0.
\end{equation}
A detailed quantitative characterization of classical and quantum
random ciphers is available in \cite{nair06}.

By \textit{standard cryptography}, we shall mean that Eve and Bob
both observe the same ciphertext random variable, i.e., $Y^{\rm
E}_{n}=Y_{n}^{\rm B}=Y_{n}$. Note that in such a standard cipher,
random or nonrandom, the following \emph{Shannon limit}
\cite{massey88,shannon49} applies:
\begin{equation} \label{shannonlimit} H(X_n|Y_n) \leq H(K).
\end{equation}
By \textit{information-theoretic security} on the data, we mean
that Eve cannot pin down uniquely the plaintext from the
ciphertext, i.e.,
\begin{equation}\label{ITsecurity}
H(X_n|Y_n)\neq 0.
\end{equation}
The \emph{level} of such security is quantified by $H(X_n|Y_n)$.
Shannon has defined \textit{perfect security} \cite{shannon49} to
mean that the plaintext is statistically independent of the
ciphertext, i.e.,
\begin{equation} H(X_n|Y_n)=H(X_n).
\end{equation}
We shall use \emph{near-perfect} security to mean $H(X_n|Y_n)
\sim H(X_n).$  Security statements on ciphers are naturally made
with respect to particular possible attacks. We will discuss the
usual cases of ciphertext-only attack, known-plaintext attack, and
statistical attack in the next subsection. We now turn to key
generation.

\subsection{Key Generation}
The objective of key generation is to generate fresh keys. By a
\emph{fresh} key, we mean a random variable $K^g$ shared by the
users from processing on $X_n$ for which
\begin{equation} \label{freshkey} H(K^g|KY_n^E) \sim H(K^g)
\end{equation} for \emph{some} $n$. Here $K$ is any secret key
used in the key generation protocol. In other words, one needs to
generate \emph{additional} randomness statistically independent of
previous shared randomness such as a secret key used in the
protocol. The two major approaches to key generation are via
classical noise \cite{maurer93} and BB84-type \cite{gisin02}
quantum cryptography. With the advent of quantum cryptography, the
term `unconditional security' has come to be used, unfortunately,
in many possible senses. By \emph{unconditional security}, we
shall mean near-perfect information-theoretic security against all
attacks consistent with the known laws of quantum physics.

Using Eq.~(\ref{decryption}), it is easily seen that, in standard
cryptography, $X_n$, or any publicly announced function thereof,
cannot serve as fresh key. This is because all the uncertainty in
$X_{n}$ is derived from $K$, however long $n$ is, and therefore
$H(K^g|KY_n)=0$.

While key generation is impossible in standard cryptography, it
becomes possible in principle in a situation where $Y_n^E \neq
Y_n^B.$ This necessary condition must be supplemented by a
condition for \textit{advantage creation} \cite{yuen04}, e.g.,
\begin{equation} \label{keygeneration}
H(X_n|KY_n^E)>H(X_n|KY_n^B).
\end{equation}
In (\ref{keygeneration}), the key $K$ is conceptually granted to
Eve after her measurements to bound the information she may
possibly obtain by any collective classical processing that takes
advantage of the correlations introduced by $K$. We mention here
that even when there is no \textit{a priori} advantage, provided
$Y_n^B \neq Y_n^E$, advantage may often be created by
\textit{advantage distillation}, as e.g., through post-detection
selection so that Eq.(\ref{keygeneration}) is satisfied for the
selected results. Keyed Communication in Quantum Noise, called KCQ
in \cite{yuen04} and hereafter, provides one way of creating
advantage for fresh key generation from the performance difference
between the optimal quantum receivers designed with and without
knowledge of the secret key. Some of the advantages of such an
approach to key generation would be indicated later, and further
details can be found in \cite{yuen04,yuen06}.

Even when information-theoretic security does not obtain, so that
the data or the key is in fact uniquely determined by the
ciphertext (we shall see in Subsection 2.4 that this is the usual
situation in standard cryptography when the plaintext has known
nonuniform statistics), we may still speak of
\textit{complexity-based} security. This refers to the amount of
computation or resources required to find the unique plaintext
$X_n$ or key $K$ corresponding to the observed $Y_n$. In practice,
forcing a large amount of computation on Eve can provide very
effective security. In fact, standard ciphers owe their widespread
use to the absence of known efficient algorithms that can find the
unique key or plaintext from the ciphertext, with or without some
known plaintext. Note that the security of a system is especially
good if the complexity goes exponentially in $|K|$, resulting in a
search problem that \textit{cannot} be efficiently handled even by
a quantum computer. In contrast to merely `hard' problems such as
factoring integers or even NP-complete problems, for which
complexity is not quantified, \emph{exponential complexity} is a
guarantee of realistic security \emph{as good as unconditional
security}. This is because a quantity that is exponential in a
system parameter can easily become so large as to be impossible to
achieve. For example, it is a fact as certain as any physical law
that one cannot have $10^{600}$ beamsplitters (See our response to
the first attack of Lo and Ko in Section 4.) on the earth, or in
the whole known universe for that matter
--- this can be seen merely from size considerations. Similar
remarks hold for exponential computing time requirements. However,
neither $\alpha\eta$ nor any standard cipher has been proven to
require exponential resources to break.

\subsection{Classes of attacks in quantum cryptography}

In our KCQ approach, we conceptually grant a copy of the
transmitted state to Eve for the purpose of bounding her
information. Thus, there is no need of considering what kind of
probe she uses. For further details, see \cite{yuen04,yuen06}.
Accordingly, we will classify attacks a little differently from
the usual case in BB84 protocols, basing our classification only
on the quantum measurement or processing Eve may make.

By an \textit{individual attack}, we mean one where the same
measurement is made in every qubit/qumode and the results are
processed independently of one another. Obviously, the latter is
an artificial and unrealistic constraint on an attack, but
analyses under this assumption are standard for BB84. In this
connection, we note that in the BB84 literature, one often finds
individual attacks being defined only by Eve's qubit-by-qubit
probes and measurements, but with the actual analysis of such
attacks being carried out with the \textit{further assumption}
that no classical collective processing is used, so that Eve has
independent, identically distributed (iid) random variables on her
bit estimates. This assumption renders the results rather
meaningless, as Eve can easily jointly process the quantum
measurement results to take advantage of the considerable
\textit{side information} available to her from announcements on
the classical public channel. It is a subtle task to properly
include such side information in the security proofs of BB84-type
protocols, one that we will elaborate upon in future papers.
However, it is this definition of individual attack that has been
used for our information-theoretic security claims in \cite{prl}.

By a \textit{collective attack}, we mean one where the same
measurement is made in every qubit/qumode but where joint
classical processing of the results is allowed. Conceptually, one
may also consider the most general attacks on classical systems to
be in this class.  We will refer to a particular collective attack
on $\alpha\eta$ using heterodyne or phase measurement on each
qumode later in this paper. Note also that encryption of a known
plaintext with all possible keys followed by comparison of the
result to the observed mode-by-mode measurement result $Y_n^E$
(i.e. a \textit{brute-force search}) is a collective attack, since
the correlations between the ciphertext symbols introduced during
encryption are being used. Note that our use of the term
``collective attack'' is different from the BB84 case, due to the
fact that there is no need to account for probe setting in our KCQ
approach. Finally, for us, a \textit{joint attack} refers to one
where a joint quantum measurement on the entire sequence of
qubits/qumodes is allowed. This is the most general attack in the
present circumstance, and must be allowed in any claim of
unconditional security.

\subsection{Security against known-plaintext attacks and statistical attacks}
In this subsection, we describe some results in classical
cryptography that are not readily available in the literature. For
a standard cipher, the conditional entropy $H(X_n|Y_n)$ describes
the level of information-theoretic security of the data $X_n$, and
$H(K|Y_n)$ describes the information-theoretic security of the
key. The attacks considered in cryptography are ciphertext-only
attacks, and known-plaintext or chosen-plaintext attacks. There is
in the literature an ambiguity in the term `ciphertext-only
attack' regarding whether the \emph{a priori} probability
distribution $p(X_n)$ of the data is considered known to the
attacker or is completely random to her.  To avoid confusion, we
will use the term \textit{ciphertext-only attack} to refer to the
case where $p(X_n)$ is completely random to Eve,
\textit{statistical attack} to refer to the case when some
information on $X_n$ in the form of a nonuniform $p(X_n)$ is
available to Eve, \textit{known-plaintext attack} to refer to the
case when some specific $X_n$ is known to Eve, and
\textit{chosen-plaintext attack} to refer to the case when some
specific $X_n$ is chosen by Eve. Generally, our results referring
to known-plaintext attacks are valid in their qualitative
conclusions also for chosen-plaintext attacks. (Note that we are
restricting ourselves to private-key cryptography -- This is not
generally true in public-key cryptography.) Therefore, our use of
the term `known-plaintext attack' may be taken to include
chosen-plaintext attacks also, for symmetric-key direct
encryption.

In standard cryptography, one typically does not worry about
ciphertext-only attack on nonrandom ciphers, for which
Eq.~(\ref{shannonlimit}) is satisfied with equality for large $n$
for the designed key length $ |K|=H(K)$ under some `nondegeneracy'
condition \cite{jkm}. In such situations, it is also the case that
$H(K|Y_n)=H(K)$ so that no attack on the key is possible
\cite{jkm}. However, under statistical and known-plaintext
attacks, this is no longer the case and Eve can launch an attack
on the key and use her resulting information on the key to get at
future data. Indeed, it is such attacks that are the focus of
concern in standard ciphers such as the Advanced Encryption
Standard (AES). For statistical attacks, Shannon \cite{shannon49}
characterized the security by the {\em unicity distance} $n_0$
\emph{(for statistical attacks)}, which is defined to be the input
data length at which $H(K|Y_{n_{0}}) = 0$. For a nonrandom cipher
defined by (\ref{nonrandom}), he derived an estimate on $n_0$ that
is \emph{independent} of the cipher in terms of the data entropy.
This estimate is, unfortunately, not a rigorous bound. Indeed, one
of the inequalities in the chain goes in the wrong direction in
the derivation, although it works well empirically for English
where $n_0 \sim 25$ characters. Generally, it is easy to see that
a finite unicity distance exists only if, for some $n$, there is
no \emph{redundant key use} in the cryptosystem, i.e., no
plaintext sequence $X_n$ is mapped to the same ciphertext $Y_n$ by
more than one possible key value. With redundant key use, one
cannot pin down the key but it seems one also could not enhance
the system security either, and so is merely wasteful. The exact
possibilities will be analyzed elsewhere. A nonrandom cipher is
called \emph{nondegenerate} in this paper if it has no redundant
key use either at some finite $n$ or for $n \rightarrow \infty$. A
\emph{random} cipher will be called nondegenerate when each of its
nonrandom reductions is nondegenerate (See \cite{nair06}). Under
the condition
\begin{equation} \label{nondegenerate}
\lim_{n \rightarrow \infty} H(Y_n|X_n) = H(K),\end{equation} which
is similar but not identical to the definition of a
`nondegenerate' cipher given in \cite{jkm}, one may show that,
when (\ref{nonrandom}) holds, one has
\begin{equation} \label{broken} \lim_{n \rightarrow \infty}
H(K|X_nY_n) = 0.
\end{equation}
In general, for a nonrandom cipher, we define a
 \emph{nondegeneracy distance} $n_d$ to be the smallest $n$ such that
 \begin{equation} \label{nondegdist}
 H(Y_{n}|X_{n})=H(K)
 \end{equation}
 holds, with $n_d =\infty$ if (\ref{nondegenerate}) holds and there is no finite $n$ satisfying
 (\ref{nondegdist}). Thus, a nonrandom cipher is nondegenerate in
 our sense if it has a nondegeneracy distance, finite or infinite. In general, of course, the cipher
 may be \emph{degenerate}, i.e., it has no nondegeneracy distance. We have the result given by
 Proposition A of Appendix A that, \emph{under
known-plaintext attack, a nonrandom nondegenerate cipher is broken
at data length $n=n_d$}. This is also the minimum length of data
needed to break the cipher for any possible known-plaintext $X_n$.
Many ciphers including the one-time pad and LFSRs (linear feedback
shift registers \cite{stinson}) have finite $n_d$. For
chosen-plaintext attacks, the above definitions and results apply
when the random variable $X_n$ is replaced by a specific
$X_n=x_n$.

The above result has not been given in the literature, perhaps
because $H(K|X_nY_n)$ has not been used previously to characterize
known-plaintext attacks. But it is assumed to be true in
cryptography practice that $K$ would be pinned down for
sufficiently long $n$ in a nonrandom `nondegenerate' cipher.
However, there is \emph{no} analogous result on random ciphers,
since under randomization Eq. (\ref{nonrandom}), and usually
(\ref{nondegdist}) also, does not hold for any $n$.

The following result is similar to one in \cite{jkm,gunther88}.
The homophonic substitution algorithms provided in these
references work also for finite sequences, and may result in data
compression rather than data expansion depending on the plaintext.
\newline \newline \emph{Proposition B}
\newline  \newline In a statistical attack on nonuniform iid $X_n$, homophonic
substitution randomization \cite{jkm,gunther88} on a nonrandom
nondegenerate cipher can be used to convert the attack into a
ciphertext-only one, thus completely protecting the key. \newline
\newline This reduction does not work for known-plaintext attacks.
The problem of attacking a symmetric-key random cipher has
received limited attention because they are not used in practice
due to the associated reduction in effective bandwidth or data
rate, and also due to the uncertainty on the actual input
statistics needed for homophonic substitution randomization. Thus,
the quantitative security of random ciphers against
known-plaintext attacks is not known theoretically or empirically,
although in principle random ciphers could defeat statistical
attacks according to Proposition B. All that is clear is that
random ciphers are harder to break than the corresponding
nonrandom ones, because a given pair $(X_n,Y_n)$ may arise from
more possible keys due to the randomization. See ref. [4] for a
detailed elucidation.

If a random cipher is nondegenerate, we say it has
\emph{information-theoretic security against known-plaintext
attacks} when
\begin{equation}  \label{ITsecurityKPT}
\inf_n H(K|X_nY_n) > 0,
\end{equation}
i.e., if $H(K|X_nY_n)$ cannot be made arbitrarily small whatever
$n$ is. The actual level of the information-theoretic security is
quantified by the left side of (\ref{ITsecurityKPT}). As in the
nonrandom case, only for a nondegenerate cipher, i.e., one with no
redundant key use, is it \emph{meaningful} to measure key security
with entropy. It is \emph{possible} that some random ciphers
possess such information-theoretic security. See Appendix A.

We define the \emph{unicity distance $n_1$ for known-plaintext
attacks}, for both nondegenerate random and nonrandom ciphers, as
the smallest $n$, if it exists, for which
\begin{equation}\label{n1} H(K|X_nY_n)=0.\end{equation} The
unicity distance $n_1$ is defined to be infinity if (\ref{n1})
holds for $n \rightarrow \infty$. Any cipher with
information-theoretic security against known-plaintext attacks has
no unicity distance $n_1$. For a nondegenerate nonrandom cipher,
we have shown in Appendix A that $n_1=n_d$. We shall see in the
next section that $\alpha\eta$ can be considered a random cipher
in the above sense under collective attacks, but with no reduction
in effective data rate. (Recall that collective attacks are the
most general in classical ciphers.) Thus, the statement in
\cite{loko} that ``known-plaintext attacks are rather standard and
were successfully launched against both the Germans and the
Japanese in World War II'' is an oversimplification, since the
ciphers referred to in it were nonrandom.

\section{$\alpha \eta$ Direct Encryption and Key Generation}
Consider the original experimental scheme $\alpha\eta$ (called
Y-00 in Japan) as described in [2] and depicted in
Fig.~1. Alice encodes each data bit into a coherent
state in a \textit{qumode}, i.e., an infinite-dimensional Hilbert
space (the terminology is analogous to the use of \textit{qubit}
for a two-dimensional Hilbert space), of the form (we use a single
qumode representation rather than a two-qumode one for
illustration)
\begin{equation} \label{states}
    |\alpha_{\ell}\rangle=|\alpha_{0}(\cos\theta_{\ell}+i
    \sin\theta_{\ell})\rangle
\end{equation}
where $\alpha_{0}$ is real, $\theta_{\ell}=2\pi\ell/M$, and $\ell
\in \{0,..., M-1\}$. The $M$ states are divided into $M/2$ basis
pairs of antipodal signals $\{|\pm \alpha_{\ell} \rangle \}$ with
$-\alpha_{\ell}=\alpha_{\ell+M/2}$.  A seed key $K$ of bit length
$|K|$ is used to drive a conventional encryption mechanism whose
output is a much longer running key $K^{\prime}$ that is used to
determine, for each qumode carrying the bit $b \{=0,1\}$, which
pair $\{|\pm\alpha_{\ell}\rangle\}$ is to be used. The bit $b$
could either be part of the plaintext in a direct encryption
system (as is the case in \cite{prl}) or it could be a raw key bit
from a random number generator. Bob utilizes a quantum receiver to
decide on $b$ knowing which particular pair
$\{|\pm\alpha_{\ell}\rangle\}$ is to be discriminated. On the
other hand, Eve needs to pick a quantum measurement for her attack
in the absence of the basis knowledge provided by the seed or
running key. The difference in their resulting receiver
performances is a quantum effect
that constitutes the ground, as we shall see in subsequent
subsections, both for making $\alpha\eta$ a random cipher for
direct encryption, and for possible advantage creation vis-a-vis
key generation. To avoid confusion, we shall use the term
`$\alpha\eta$' to refer only to the direct encryption system
following our practice in \cite{prl}. When we want to use the same
system as part of a key generation protocol, we shall refer to it
as `$\alpha\eta$-Key Generation' or `$\alpha\eta$-KG'. We discuss
$\alpha\eta$ and $\alpha\eta$-KG in turn in the next two
subsections.

Note that since the
quantum-measurement noise is irreducible, such advantage creation
may result in an unconditionally secure key-generation protocol.
In contrast, in a classical situation including noise, the
simultaneous measurement of the amplitude and phase of the signal,
as realized by heterodyning, provides the general optimal
measurement for both Bob and Eve; thus preventing any advantage
creation under our approach that grants Eve a copy of the state
for the purpose of bounding her information. We may remark that
since a discrete quantum measurement is employed by the users,
$\alpha\eta$ and $\alpha\eta$-KG are \emph{not}
continuous-variable quantum cryptosystems. In particular, their
security is not directly derived from any uncertainty relation for
observables with either continuous or discrete spectrum.
\begin{figure*}\begin{center}
\includegraphics[width=13cm]{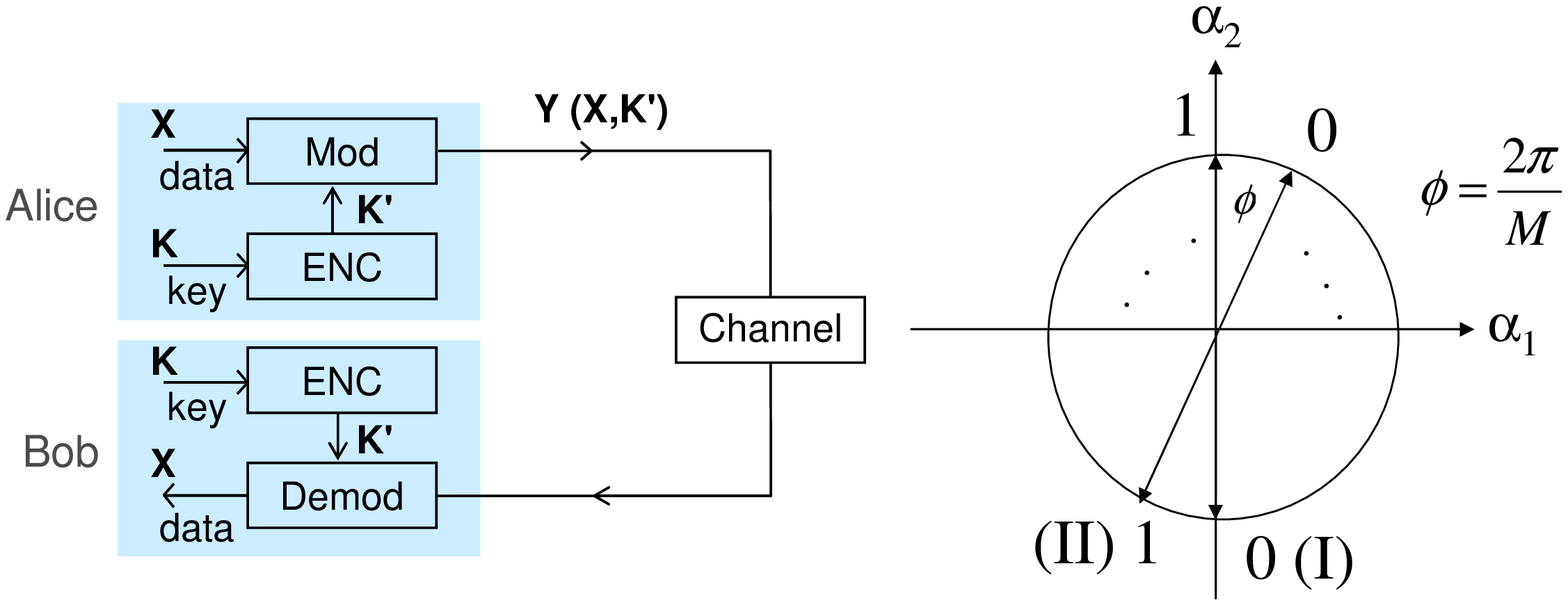}
 \fcaption{Left:Overall schematic of the $\alpha\eta$ scheme. Right: Depiction of
$M/2$ bases with interleaved logical state mappings.}\label{setup}
\end{center}\end{figure*}

\subsection{$\alpha\eta$ Direct Encryption}

Let $X_{n},Y_{n}^{\rm E},Y_{n}^{\rm B}$ be the classical random
vectors describing respectively the data, Eve's observation, and
Bob's observation. Eve may make any quantum measurement on her
copy of the quantum signal to obtain $Y^{\rm E}_{n}$ in her
attack. One then considers the error in her estimation of $X_n$.
As an example, consider the attack where Eve makes a heterodyne
measurement or a phase measurement on each qumode
\cite{yuen04,pla05}. Under such an attack, $\alpha\eta$ becomes
essentially a classical random cipher (in the sense of Section 2),
because it satisfies
\begin{equation} \label{eve} H(X_n|Y_n^E, K) \sim 0
\end{equation}
along with Eq.~(\ref{random}) for the experimental parameters of
\cite{prl,optlett03,pra05,ptl05}. Under Eq.~(\ref{eve}),
Eq.~(\ref{shannonlimit}) also obtains and the data security is no
better than $|K|$ as in all standard symmetric key ciphers. Still,
heterodyning by Eve does not reduce $\alpha \eta$ to a classical
nonrandom stream cipher, as claimed in \cite{nishioka}. Rather, it
becomes a \textit{random cipher} as already pointed out in
\cite{yuen04}. For each transmitted qumode, the plaintext alphabet
is $\{0,1\}$ and the ciphertext alphabet is any point on the
circle of Fig. 1 when a phase measurement is made by Eve, and is
any point in the plane when a heterodyne measurement is made. Note
that the ciphertext \emph{alphabet} depends on what quantum
measurement is made by the attacker. However, it can at most be
reduced to an $M$-ary one by collapsing the continuous outcomes
into $M$ disjoint sets. This is so because such an alphabet is the
smallest possible ciphertext alphabet such that it is possible to
decrypt for \emph{every} possible value of ciphertext and key. We
have elaborated on this point in Section 5 of \cite{nair06}.
Hence, $\alpha\eta$ is a random cipher against attacks on the key,
and cannot be reduced to an additive stream cipher, which is
nonrandom. When it is
forced to become nonrandom, even just for Bob, it becomes noisy.
See our reply \cite{pla05} to the attack in \cite{nishioka} for
more details. Also see their subsequent response \cite{nishioka2}
based on a confusion regarding the interpretation of
Eq.(\ref{eve}), which is valid for our $\alpha\eta$ system of
\cite{prl}. Further elaboration is available in \cite{nair06}.

Observe that the randomization in $\alpha\eta$ can be accomplished classically in
principle, but not in current practice. This is because true
random numbers can only be generated physically, not by an
algorithm, and the practical rate for such generation is many
orders of magnitude below the $\sim$ Gbps rate in our experiments
where the coherent-state quantum noise does the randomization
\textit{automatically}. Furthermore, our physical ``analog''
scheme does not sacrifice bandwidth or data rate compared to other
known randomization techniques. This is because Bob resolves only two, not $M$ possibilities.
Another important point with regard to
physical cryptosystems like $\alpha\eta$, whether random or
nonrandom, is that they require the attacker to make \emph{analog}
or at least $M$-ary observations, i.e., to attack the system at
the \emph{physical} level,  even though the data transmitted is
binary. In particular, as indicated above, it is impossible to
launch a known-plaintext attack on the key using just the binary
output, available for instance at a computer terminal.

While the original $\alpha\eta$ scheme of Fig.~1 is a random
cipher under collective attacks made without knowledge of the key
$K$, or more generally, under qumode-by-qumode measurements that
can vary from qumode to qumode, it is still a nonrandom cipher in
the sense of quantum states. See also ref. \cite{nair06}. The
technique called Deliberate Signal Randomization (DSR) described
in \cite{yuen04} would make it a random cipher even with respect
to quantum states. This amounts to randomizing (privately in the
sense of \cite{massey88}) the state transmitted so as to cover a
half-circle around the basis chosen by the running key. The
security of such ciphers is an open area of research. While we
will not delve into the details of DSR in this paper, it may be
mentioned that at the mesoscopic signal levels used in
\cite{prl,optlett03,pra05,ptl05}, DSR with an error-correcting
code on top may be expected to induce many errors for Eve while
Bob remains essentially error-free. The reason is similar to that
for Eq.~(4) in Ref. \cite{pla05}, with advantage for Bob due to
the optimal receiver performance difference described in the next
subsection and in \cite{yuen04}. Thus, information-theoretic
security is expected \cite{yuen04} for the key, and at a level far
exceeding the Shannon limit for the data, when DSR is employed on
$\alpha\eta$. Instead of DSR, a keyed `mapper' that varies the
mapping from the running key to the basis from qumode to qumode
can also be employed, including perhaps a polarity (0 or 1) bit to
enhance security. Even with the original $\alpha\eta$, it can be
expected that the randomization or coherent-state noise would
increase the unicity distance $n_1$ compared to the ENC box alone
used as a cipher. Further details can be found in \cite{nair06}.

For the direct-encryption experiments in Refs.
\cite{prl,optlett03,pra05,ptl05}, we have claimed
``unconditional'' security only against ciphertext-only individual
attacks. We have claimed only \emph{exponential complexity-based}
security against assisted brute-force search (See \cite{nair06})
known-plaintext attacks, which is more than the security provided
just by the ENC box of Fig.1 \cite{pla05}. However,
information-theoretic security, even at the near-perfect level for
both the key and the data, is possible with additional techniques
or CPPM-type schemes described in \cite{yuen04}. Detailed
treatment will be given in the future. But see also ref.
\cite{nair06}.

We summarize the main known advantages of $\alpha\eta$ compared to
previous ciphers:
\begin{enumerate} [(1)]
\item It has more assisted brute-force search complexity for attacks on the
key compared to the case when the quantum noise is turned off. For
an explicit claim, see \cite{nair06}.

\item It may, especially when supplemented with further techniques, have information-theoretic security against
known-plaintext attacks that is not possible with nonrandom
ciphers.

\item With added Deliberate Signal Randomization (DSR), it is expected to have information-theoretic
security on the data far exceeding the Shannon limit.

\item It has high-speed private true randomization (from quantum noise that even Alice does not know), which is not
possible otherwise with current or foreseeable technology.

\item It suffers no reduction in data rate compared to other known
random ciphers.

\item The key cannot be successfully attacked from a computer terminal
with bit outputs, as is possible with standard ciphers.
\end{enumerate}

\subsection{$\alpha\eta$ Key Generation}
One needs to clearly distinguish the use of such a scheme for key
generation versus data encryption.  It may first appear that if
the system is secure for data encryption, it would also be secure
for key generation if the transmitted data are subsequently used
as new key. It seems to be the view taken in
\cite{loko,nishioka,ys05} that we have made such a claim, which we
have not. The situation may be delineated as follows. Following
the notations of the last subsection, Eve may make any quantum
measurement on her copy of the quantum signal to obtain $Y^{\rm
E}_{n}$ in her attack. Such a measurement is made \textit{without
the knowledge of} $K$. It is then used together with the value of
$K$ to estimate the data $X_n$. Although Eve is not actually given
$K$ after her measurements, we give it to her \emph{conceptually}
for the purpose of bounding her information. The conditions for
unconditional security are complicated, and to satisfy them one
needs to extend $\alpha \eta$-KG in different possible ways, such
as DSR and CPPM described in \cite{yuen04}. However, against
attacks with a fixed qumode measurement, Eq.~(\ref{keygeneration})
is sufficient and can be readily seen to hold as follows.

With $S\equiv|\alpha_0|^2$ being the average photon number in the
states (11), the bit-error rate for Bob with the optimum quantum
receiver \cite{helstrom76} is
\begin{equation}
\label{optimum} P_{b}=\frac{1}{4}e^{-4S}.
\end{equation}
The bit-error rate for heterodyning, considered as a possible
attack, is the well-known Gaussian result \begin{equation}
\label{heterodyne} P_{b}^{\rm
het}\sim\frac{1}{2}e^{-S},\end{equation} and that for the
optimum-phase measurement tailored to the states in (\ref{states})
is
\begin{equation}\label{phase} P_{b}^{\rm ph}\sim\frac{1}{2}e^{-2S}
\end{equation}over a wide range of $S$.  The difference between Eq.~(16) and
Eq.~(17-18) allows key generation at any value of $S$ if $n$ is
long enough. With a mesoscopic signal level $S\sim7$ photons, one
has $P_{b}\sim 10^{-12}$, $P_{b}^{\rm het}\sim 10^{-3}$, and
$P_{b}^{\rm ph}\sim 10^{-6}$. If the data arrives at a rate of 1
Gbps, Bob is likely to have $10^{9}$ error-free bits in 1 second,
while Eve would have at least (recall that she actually does not
have the key even after her measurements) $\sim 10^{6}$ or $\sim
10^{3}$ errors in her $10^{9}$ bits with heterodyne or the
optimum-phase measurement (which has no known experimental
realization).  With the usual privacy amplification \cite{bbcm95},
the users can then generate $\sim 10^{6}$ or $\sim 10^{3}$  bits
in a 1 second interval by eliminating Eve's information. While
these parameter values are not particularly remarkable due to the
loose bound and have not been experimentally demonstrated, they
illustrate the new KCQ principle of quantum key generation
introduced in \cite{yuen04} that creates advantage via the
difference between optimal quantum receiver performance with
versus without knowledge of a secret key, which is more powerful
than the previous BB84 principle since it does not rely on
intrusion-level estimation to create advantage. Also note that due
to the 3 dB advantage limitation of binary signaling (compare Eq.
(\ref{phase})and Eq.(\ref{optimum})), one may use the CPPM scheme
\cite{yuen04} and its extensions instead of $\alpha\eta$-KG for
key generation over long distances. Within the confines of binary
signaling, the throughput, though not the advantage, can be
greatly increased even for large $S$ by moving the state close to
the decision boundary. Detailed treatments will be given in the
future.

The heterodyne attack on $\alpha \eta$ discussed above can of
course be launched also on an $\alpha\eta$ Key Generation system.
For parameter values, i.e., values of $S$, $M$ and $n$, such that
Eq. (\ref{eve}) holds, key generation with information-theoretic
security is impossible in principle, since the Shannon limit
(\ref{shannonlimit}) holds. This point is \textit{missed} in all
the criticisms of $\alpha \eta$ Key Generation
\cite{loko,nishioka,ys05}, but was explicitly stated in the first
version of Ref. \cite{yuen04}. It is at least implicit in Ref.
\cite{prl} where we said the experiment has to be modified for key
generation, and also mentioned the KCQ Key Generation Principle of
optimal quantum receiver performance difference. One simple way to
break the Shannon limit (\ref{shannonlimit}) and protect the key
at the same time, is to employ DSR. As noted in Section 3.1, its use in $\alpha\eta$
direct encryption is expected to provide information-theoretic
security for the key and at a level far exceeding the limit
(\ref{shannonlimit}) for the data. We mention these possible
approaches to make it clear that we were aware of the limitations
of $\alpha\eta$  and that we need additional techniques to obtain
unconditinal security.

\section{The Lo-Ko Attacks}
\subsection{Review of Attacks in \cite{loko}}
Ref. [1] first describes a known-plaintext attack on the original
$\alpha \eta$ of \cite{prl} that can be launched when the channel
loss allows Eve to have $2^{|K|}$ copies of the states Bob would
receive. With $2^{|K|}$ copies, it is claimed that Eve can use
each possible seed key to implement a decryption system similar to
Bob's, and by comparing the outputs to the known-plaintext of some
\emph{unspecified} length $s$, can determine the key. Eve thus
needs only beamsplitters and detectors similar to Bob's to
undermine the system. We shall call this attack \emph{Attack I} in
the sequel. A variant of this attack is also described, in which
Eve is assumed to know $r$ $s-$bit sequences of plaintext, where
$r(1-\eta) \geq 2^{|K|} \eta$. In other words, the channel
transmittance $\eta$ is such that Eve has in her possession,
including repeated copies, $2^{|K|}$ ciphertext-states, each
corresponding to a known $s-$bit sequence. What $s$ needs to be is
again unspecified. It is claimed that an exhaustive trial of keys
would again pin down the key in this case. These attacks are also
claimed to work, without \emph{any} supporting argument, when the
plaintext is not exactly known, but is drawn from a language,
e.g., English.

It is further argued that even in just 3 dB loss (which is not
required under our approach of granting Eve a copy of the quantum
signal), a Grover quantum search (that will be called \emph{Attack
II}) would succeed in finding $K$ under a known-plaintext attack
when $n=\infty$, because then there is only a single possible key
value that would give rise to the overall ciphertext-state from
the known data $X_n$. This latter claim is in turn justified by
the ``asymptotic orthogonality'' of the ciphertext-states
corresponding to different key values, although exactly how this
asymptotic orthogonality occurs for different choices of the ENC
box in Fig.1, including the LFSR used, is not described.  The
purpose of this argument is presumably to claim that a limiting
statement such as (\ref{broken})
must be true, thus undermining the system under a known-plaintext
attack \textit{for large enough} $n$. When the plaintext $X_n$ is
not exactly known but is not completely random, i.e., under a
statistical attack, such a result is also claimed to hold
\textit{without} any argument. Also, no estimate of the
convergence rate in $n$ is provided for either asymptotic
orthogonality or for Eq.(\ref{broken}).

Ref. [1] then assumes that $\alpha \eta$ Key Generation, in which
$X_n$ is taken to be completely random as in all key-generation
protocols (so that there is no possibility of a known-plaintext or
statistical attack of any kind, at least before the generated key
is used in another cipher), proceeds by utilizing the output bits
$Y_n = X_n$ directly as key bits to XOR or ``one-time pad'' on new
data. With known-plaintext attack on these new data, the $X_n$
would be known and the previously described known-plaintext
attacks I and II can be applied on the ciphertext-states to find
$K$.

\subsection{Response to Attacks}
We will first respond to these attacks for direct encryption. The
first gap in Attack I is that the length of known-plaintext $n_1$
needed to uniquely fix the key is not specified. From Subsection
2.3, we see that Eve needs length equal to the nondegeneracy
distance $n_d$ (\ref{nondegdist}) of the ENC box of Fig.1
 to fix the key from exact input-output pairs of the ENC box alone.
 Actually, $s=n_1$ needs to be larger than this nondegeneracy
 distance $n_d$ due to the quantum noise randomization. Note also that the ENC box could be chosen to be degenerate,
so that it does not even have a nondegeneracy distance and the key
could never be pinned down. However, since the LFSR used in
\cite{prl} is actually nondegenerate, we will not dwell on this
point. As it stands, the attack is seriously incomplete without
specifying what $s=n_1$ is or at least providing estimates of it.
This corresponds to defect One in our Introduction.

Furthermore, Attack I requires the product $r(1-\eta)$ to be
bigger than $\eta 2^{|K|}$, which implies either $r$ or $1/\eta$
is at least exponential in $|K|/2$. Thus, Attack I can be thwarted
by increasing the key length linearly, which is relatively easy.
As an example, for the key length $|K| \sim 2\times 10^3$ used in
\cite{prl}, one needs a loss of $6\times 10^3$ dB for $r=1$, which
corresponds to propagation over $\sim 3\times 10^4$ km in the best
available fiber, which has a loss of 0.2 dB/km. No conceivable
one-stage communication line can be expected to operate over such
a long distance. Any future improvements in the loss figure of
fibers can only make Eve's task harder because the number of
copies she can tap decreases along with the loss.

If the exponential loss requirement is replaced by that of an
exponential length of data, it is equally fanciful. For the key
length $|K| \sim 2\times 10^3$, $r=2^{|K|}$ corresponds to $\sim
10^{600}$ bits of data. How could Eve input $\sim 10^{600}$ bits
of data in a chosen-plaintext attack, or know $\sim 10^{600}$ bits
in a known-plaintext attack? In any case, even if such large loss
obtains, the attacker still has the problem of requiring an
\textit{exponential number} of devices (beamsplitters \emph{and}
detectors in this case) and doing an exponential amount of
processing. Apart from size and time limitations mentioned in
Section 2, it seems not possible to ever get $\sim 10^{600}$
devices corresponding to the above key length, considering that
the total number of elementary particles in the universe is less
than $ 10^{100}$. This corresponds to defect Two in the
Introduction. We should also mention that $\alpha\eta$ was claimed
in \cite{prl} to be proved secure against known-plaintext attacks
only in the brute-force search sense and not
information-theoretically, and so the above attacks do not
contradict any claim in \cite{prl} even if they were successful.

Before proceeding to Attack II, we first distinguish the following
\emph{four} distinct kinds of statements that can be made on a
quantity $\epsilon(n)$, basing roughly on the value of $n$ being
considered:
\begin{enumerate}[(i)]
\item The value of $\epsilon(n)$ at a finite $n$. This is of interest
for a realistic implementation ---
typically $n \sim 10^2-10^4$ is the limit for joint processing of
a single block.

\item The case expressed by a limit statement on some
quantity of interest $\epsilon(n) \rightarrow 0$ \textit{with}
quantitative convergence rate estimate $0 \leq \epsilon(n) \leq
f(n)$ for $n \geq N$ and some large enough $N$ and a known
function $f(n) \rightarrow 0$.

\item  The case of the limit statement $\lim_{n \rightarrow \infty} \epsilon(n) =0$
\textit{without} convergence rate estimate. Thus, it is not known
how large $n$ needs to be for $\epsilon(n)$ to be below a certain
given level $\epsilon_0$.

\item The case of the value $\epsilon(\infty)$ \textit{at} $\infty$. Note that
the limiting value of $\epsilon(n)$ in Case (iii) above may be
different from $\epsilon(\infty)$ due to failure of continuity at
$n=\infty$.
\end{enumerate}

Observe that the statements in Cases (i)-(iii) are, in that order,
progressively weaker statements on the quantity of interest. Case
(iv), however, is \textit{independent} of the previous cases, and
can be asserted by evaluating $\epsilon(\infty)$ by a route that
does not even require $\epsilon(n)$ at finite $n$. In turn,
knowing $\epsilon(\infty)$ does not allow one to make even a limit
statement of the form of Case (iii) unless one can prove
continuity at $n=\infty$. We have classified the above cases in
order to delineate exactly what Lo and Ko can claim for their
Attack II.

Let us now consider Attack II. The first obvious problem with the
argument is that Eve does not need to attack the system if she
already knows the entire $n \rightarrow \infty$ plaintext that
will be transmitted using the particular seed key. Lo and Ko give
\emph{no} analysis of their attack for the relevant case in which
the plaintext is partially known, i.e., for the case of a
statistical attack (this includes the case of Eve knowing a
fraction of the plaintext exactly) even in the $n \rightarrow
\infty$ situation. A little thought will show that the oracle
required in Grover search would have an implementation complexity
that increases indefinitely with $n$, making it prohibitive to
build in the $n \rightarrow \infty$ limit. In other words, the
search complexity is not simply $\sim 2^{|K|/2}$ but rather
increases with $n$ as well.  When there is more than one plaintext
possible, Lo and Ko presumably intend to apply Grover search for
each plaintext in turn. The number of such repeated applications
would obviously grow indefinitely with $n$ if Eve knows only a
fraction of plaintext. In case they intend that a single Grover
search be applied to cover all possible plaintexts, they need to
produce a specific oracle that would work for this case and
analyze its performance. The issue is more critical in actual
practice, because it typically does not happen that Eve knows a
large length of plaintext, let alone one that is arbitrarily long
in the unquantified sense of (iii) above, which is what their
attack entails. Furthermore, even if its $n$ dependence is
ignored, the $\sim 2^{|K|/2}$ complexity of the Grover's search
makes it practically impossible to launch for $|K| \sim 2 \times
10^3$. Similar to Attack I, Attack II retains all the limitations
of being exponential in the key length. This point is an instance
of the second defect mentioned in Section 1.

Our second point regarding Attack II  relates to the first general
defect described in Section 1, namely lack of rigor. We claim that
the ``asymptotic orthogonality'' in \cite{loko} is vague in that
it is not specified which sense among (ii) - (iv) is meant.
Moreover, even assuming
 that a Case (ii) statement holds, it cannot by itself be translated into even a limit
statement of the form of Eq.~(\ref{broken}). To see this, let us
assume that the pairwise inner product between any of the
$2^{|K|}$ ciphertext states $\{|\psi_k\rangle\}$ corresponding to
a known plaintext encrypted with the different keys is upper
bounded by a function $\epsilon(n)$. Let us take \begin{equation}
\label{asymportho} \lim_{n \rightarrow \infty} \epsilon(n) =0
\end{equation}
to mean ``asymptotic orthogonality'' in the sense of case (iii) or
even (ii) above. For each $n$, we can in principle calculate the
optimal probability of error $P_E(n)$ in discriminating the
$2^{|K|}$ states, which, rather than the inner product, is the
relevant quantity of \emph{operational significance}. It is clear
that $P_E(\infty)=0$ in the sense (iv) above since
$\epsilon(\infty)=0$. However, to make the limit statement
Eq.~(\ref{broken}), one needs to further show from
(\ref{asymportho}), the equivalent statement to (\ref{broken})
that
\begin{equation} \label{continuity} \lim_{n \rightarrow \infty}
P_E(n)=0,
\end{equation} perhaps from the claim that the probability of error is a
\textit{continuous} function of $\epsilon(n)$. Since the
underlying Hilbert space is expanding with $n$ and becoming
nonseparable at $n=\infty$, it is not obvious whether continuity
would hold, especially at $n=\infty$. In order to convince the
reader that the above considerations indeed have real
implications, we will in Section 5 use an asymptotic orthogonality
argument to `prove' coherent-state BB84 insecure for any non-zero
loss.

Note that \cite{loko} does not actually prove ``asymptotic
orthogonality'' in any of the senses (ii)-(iv). As discussed above
in connection with Attack I, there are conditions required on the
ENC box of Fig.1 for it to be true just in the sense (iv). On the
other hand, we believe that (\ref{asymportho}) can be proved along
with (\ref{continuity}) under proper conditions on the ENC box.
But one needs precise arguments to make clear the conditions of
validity, which \cite{loko} does not provide.

Thirdly, even if their claim is correct as a \textit{limit
statement} of the form of Eq.~(\ref{broken}), that result has no
implication in practice. Indeed, an $\alpha \eta$ system with a
LFSR for the ENC box in Fig.1 has a periodic running key output
${K}^{\prime}$ of period $n_{p}=2^{|K|}/\log_2M$. It is never
meant to be used beyond such $n _{p}$, similar to the case of
standard ciphers, even in the limit of no channel loss. A limiting
claim such as Eq.~(\ref{broken}), which falls under Case (iii)
above, does not say anything about the insecurity of the actual
system. These last two points are instances of the first and
second defects described in Section 1, namely lack of rigor and
insufficient attention to quantitative detail.

Finally, we stress that we were concerned in \cite{prl} only with
exponential-complexity based security in direct encryption
systems, which is as good as unconditional security for real
systems. Also, we may mention that various added randomization
techniques are introduced in \cite{yuen04} which would modify
$\alpha \eta$ to become a random cipher even in the sense of
quantum state.  The security of such ciphers against
known-plaintext attacks is an entirely open area of research.

We now respond to the Lo-Ko attacks on $\alpha\eta$ used as a
key-generation system. First of all, we only mentioned in
\cite{prl} the possibility in principle of using the system to do
key generation without giving a complete protocol. We did not
imply that the system for the parameters in \cite{prl} and without
any modifications would function for that purpose. Indeed, in
light of the discussion in Sections 2 and 3, the Shannon limit
(\ref{shannonlimit}) already applies to the original $\alpha \eta$
of \cite{prl} for all practical $n$. Thus, there is at most
$|K|$-bits uncertainty in $X_n$ to Eve, however long $n$ is,
leaving no possibility of key generation. Thus, Lo and Ko overlook
the fact that fresh key \emph{cannot} be generated {\em in
principle} in their use of $\alpha \eta$ for key generation.
Furthermore, even if the advantage creation condition is ignored,
\cite{loko} does not include the usual step of privacy
amplification that the users can apply to the output to make it a
shorter uniformly-random key. This omission alone already
\emph{invalidates} their argument. These two points correspond to
the defect Three of Section 1.

Since the attacks on $\alpha\eta$-KG are reduced in \cite{loko} to
attacks on $\alpha\eta$ direct encryption, they also suffer the
same problems as the attacks on direct encryption above.  Indeed,
one may conclude \textit{a priori} that the attacks in
\cite{loko}, even if successful, do not contradict the claims in
\cite{prl}, and are indeed ``outside the original design'' because
they are inapplicable in any realistic situation.

\section{Attack on coherent-state BB84}

It is claimed in \cite{loko} that ``our attacks do not apply to
BB84 or other standard QKD schemes where the quantum signals are
strictly microscopic in the sense that there is (on average) at
most one copy of the signal available.'' We will show that this is
\emph{false} by using an asymptotic orthogonality argument exactly
parallel to that in \cite{loko} which will `prove' that
coherent-state BB84 using a classical error-correction code is
insecure for \textit{any} nonzero value of loss. Although we do
not believe this latter statement to be true without
qualification, we present this argument as an example to
underscore the importance of rigorous reasoning before making the
claim that $\alpha\eta$ Key Generation is insecure under Attack II
in \cite{loko}.

We denote by $b$ the  $n$-bit string that Alice intends to
transmit to Bob in order to share a key. We denote by
$|\psi_b\rangle$ the following product coherent state used to
transmit $b$:
\begin{equation} \label{BB84states}
|\psi_b\rangle=\bigotimes_{i=1}^{i=n}
|\alpha_{b_i}^{\beta_i}\rangle.
\end{equation}
Here the superscript $\beta_i$ defines which of the two BB84 bases
is used for the $i$th transmission and the subscript $b_i$ is the
$i$th bit of $b$. The exact form of the states
$|\alpha_{b_i}^{\beta_i}\rangle$ depends on the implementation.
All that is relevant for our attack is to note the obvious fact
that, for each $\beta$, the states $|\alpha_{0}^{\beta}\rangle$
and $|\alpha_{1}^{\beta}\rangle$ are distinct, and so
$|\langle\alpha_{0}^{\beta}|\alpha_{1}^{\beta}\rangle| < 1$. The
attack works as follows: When the channel transmittance is $\eta$,
Eve simply splits a fraction $1-\eta$ of the energy using a
beamsplitter and thus has in her possession the state
\begin{equation} \label{evestate}
|\psi_b^E\rangle=\bigotimes_{i=1}^{i=n} |\sqrt{1-\eta}
\alpha_{b_i}^{\beta_i}\rangle.
\end{equation}
Eve holds this state in her quantum memory, while transmitting the
remaining energy to Bob through a lossless line. Bob is thus
totally oblivious of Eve's presence. She then listens to the
public announcements of Alice and Bob, and discards along with
Alice and Bob the bit positions where Bob observes a count in both
or no detectors. She also rotates all the component states to the
same basis according to the announcement of bases by Alice.
Accordingly, we may suppress the superscript $\beta_i$. Next,
according to the protocol, Alice and Bob estimate the error rate
on a subset of their choice. Let us now assume that $n$ refers to
the remaining subset. If the fraction of errors in the test set is
under the error threshold $\delta$, Alice and Bob select an
$(n,k,d)$ code with $d>2\delta n$ to correct the errors on the
remaining bits. After the announcement of the syndrome of $b$ with
respect to the chosen code, the number of possible $b$'s goes to
$M=2^k$ rather than the original number of possibilities $2^n$.
Eve listening on the public channel can determine which
possibilities remain, and can launch a powerful attack, as seen in
the following.

The inner product of Eve's states corresponding to two admissible
bit sequences $b$ and $b'$ is
\begin{equation} \label{ip}
|\langle\psi^E_b|\psi^E_{b'}\rangle| = \prod_{i=1}^{n}|\langle
\sqrt{1-\eta}\alpha_{b_i}|\sqrt{1-\eta}\alpha_{b'_i}\rangle|=\epsilon^{\delta_{H}(b,b')}\leq
\epsilon^d < \epsilon^{2\delta n}.
\end{equation}
Here $\delta_{H}(b,b')$ is the Hamming distance between the
strings $b$ and $b'$, which is restricted to be at least the
distance of the code $d$ and
\begin{equation}\label{epsilon}
\epsilon=|\langle
\sqrt{1-\eta}\alpha_{0}|\sqrt{1-\eta}\alpha_{1}\rangle| <
1.\end{equation}
 Since $\epsilon$ is strictly less than $1$,
Eq.~(\ref{ip}) shows that the $M$ possibilities become orthogonal
in the senses (ii)-(iv) of the previous section. One thus has the
result that the \emph{probability of error} on $b$,
$P_E(\infty)=0$ since Eve can distinguish orthogonal states
without error. If one pulls this case (iv) statement to a case
(iii) limiting statement of Eve's error probability $P_E(n)$ on
$b$, parallel to the argument in \cite{loko} which takes
(\ref{asymportho}) to (\ref{continuity}),
 it would imply that the system would become insecure for
large $n$ and any nonzero value of loss!

We stress that the above result cannot be correct for all values
of the signal energy, channel loss and other parameters such as
the code rate $k/n$, no matter how large $n$ is. One reason for
doubting its universal correctness is that it would contradict the
known classical information transmission capacity of a lossy
bosonic channel \cite{giovanetti04}. However, the \emph{line of
argument} is exactly in parallel to that of \cite{loko}.  We give
it here to demonstrate the consequences of jumping to a limiting
statement from an $n=\infty$ statement on the error probability.
However, we do believe that the above attack has not been
accounted for in the security proofs in the literature. Indeed, we
agree with [1] that it is ``interesting to study the subtle
loopholes in existing schemes,'' and that one should ``never
under-estimate the effort and ingenuity that your adversaries are
willing to spend on breaking your codes.''

\section{Other comments on Lo-Ko \cite{loko}}
In this section we comment on some other claims in [1].

\begin{enumerate}[(i)]
\item The authors of \cite{loko} seem to believe that optical
amplifiers can be and are used to compensate for an arbitrary
amount of loss provided mesoscopic signal levels are used. The
supposed existence of such high-loss links in optical systems is
perhaps the reason that they state that their ``(beamsplitter)
attack can to some extent, be implemented with current
technology.'' and that ``our attacks severely limit the extent of
such optical amplification.'' In reality, however, optical
amplifiers are noisy (i.e., degrade the probability of error for
all measurements except heterodyne \cite{yuen87}) irrespective of
signal strength. Thus, in practice, in order to retain an
acceptable signal-to-noise ratio at the output, optical amplifiers
are placed at periodic intervals in a long-distance fiber link,
with the first optical amplifier being inserted much before the
channel transmittance decays to $\sim 2^{-|K|}$ for $|K|$ in the
range of $|K|\sim 1000$. So, Lo and Ko need to specify exactly how
they would proceed to attack such an optically-amplified line for
which each section has $\eta >> 2^{-|K|}.$

Optical-amplifier noise actually provides limits of operation for
both $\alpha\eta$ and $\alpha\eta$-KG. For $\alpha\eta$-KG, the
optical amplifier noise would limit the attainable advantage
needed for key generation. For $\alpha \eta$ direct encryption,
amplifier noise is a limit when there is too much of it, but it is
a help against Eve when present in a moderate amount. See for
example \cite{ptl05}. The full story of loss and amplifiers in
$\alpha\eta$ systems depends on implementation and additional
security techniques to be deployed on top of basic $\alpha \eta$,
and is yet to be told.

\item We do not believe that their attacks can be implemented to
\emph{any} useful extent with current \emph{or} future technology,
as they require exponential resources and processing.

\item It was pointed out in [1] that ``mesoscopic states for
quantum key distribution was first proposed by Bennett and Wiesner
\cite{bennettwiesner} in 1996.'' The principle underlying that
scheme is the usual BB84 type disturbance/information tradeoff,
which is radically different from our KCQ principle. Indeed, the
mesoscopic nature of the signal in that scheme is a hindrance and
not a help on the operation of the cryptosystem due to sensitivity
problems.  This is because it is not the absolute strength of the
signals that matters, but rather whether they are
\emph{distinguishable}. The large signals in \cite{bennettwiesner}
still have only one photon average difference between them. As the
large signal gets attenuated in a lossy line, the one-photon
difference also gets attenuated correspondingly. Thus, as compared
to the small signal case, these large signals are distinguishable
at the receiver with the same absolute difficulty, but with a
bigger relative difficulty since the signal level difference is
now a much smaller percentage of the absolute level.
\end{enumerate}

\section{Conclusions}
There are two main claims in \cite{loko} against the original
$\alpha \eta$ cryptosystem of \cite{prl,yuen04}.

\begin{enumerate}[(1)]
\item It is broken in high loss channels by a beamsplitter attack
(Attack I) and in 3 dB loss by a quantum search (Attack II) when
the attacker knows a sufficiently long plaintext for Attack I and
infinitely long plaintext for Attack II. The lack of security is
taken in the information-theoretic sense that the seed key K could
be uniquely determined.

\item If the output of an $\alpha \eta$ Key Generation ($\alpha\eta$-KG) system is directly used
as the key in a ``one-time pad'' cipher, then a known-plaintext
attack on that cipher would allow one to launch the above
known-plaintext attacks on $\alpha \eta$-KG.
\end{enumerate}

Our detailed response has been given in this paper, which
describes the proper framework for discussing these attacks and shows that
the arguments in \cite{loko} fail at many levels. A brief summary
of our response follows:

For direct encryption, Attack I requires either loss that is
exponentially large in the key length or knowledge of an
exponentially long sequence of plaintext, which are both
unrealistic. Attack II, by \emph{requiring} $n$ to be infinite, is
not applicable as the original $\alpha\eta$ is designed to run for
a finite $n$. That attack also contains gaps in the reasoning for
making even a limiting insecurity statement. While it is claimed
that these attacks work when just the statistics of the plaintext
is known, it is not described how it would even proceed, not to
mention its quantitative performance. Also, the attacks, even if
successful, do not undermine our claims in \cite{prl} of
exponential assisted brute-force search complexity
under known-plaintext attacks.

The attacks on $\alpha\eta$ as a key-generation system are founded
on a key-generation protocol created by Lo and Ko, since no
key-generation system was detailed in \cite{prl}. Their
key-generation protocol omits the crucial step of privacy
amplification before the generated key is used in an encryption
system. In addition, the attacks are not relevant for the security
of $\alpha\eta$ with the parameters of \cite{prl} since, for
information-theoretic security of the key studied in \cite{loko},
the heterodyne attack described in \cite{yuen04} and this paper
and not recognized in \cite{loko} prevents the original $\alpha
\eta$ from generating fresh keys due to the Shannon limit.

Perhaps more significantly, we have described various security
features of $\alpha \eta$ that appear to be widely misunderstood,
partly because little is known on the corresponding classical or
standard cryptosystems.  We hope this paper explains our new
approach sufficiently to dispel misunderstandings, and at the same
time highlights many important considerations on quantum
cryptography in action.

\section*{Acknowledgments}
We would like to thank T.~Banwell, H.~Brandt, G.M.~D'Ariano,
M.~Foster, M.~Goodman, O.~Hirota, D.~Hughes, C.~Liang,
D.~Nicholson, M.~Ozawa, J.~Smith, P.~Toliver, and K.~Yamazaki for
many useful discussions during the continuing development of
$\alpha\eta$.

This work was supported under DARPA grant F30602-01-2-0528.

\nonumsection{References} \noindent

\section*{Appendix A -- Security Against Known-Plaintext Attack}

In this appendix, we provide a brief quantitative discussion of
known-plaintext attacks on random and nonrandom ciphers that is
not available in the literature. For nonrandom ciphers, we have
the following
\newline \newline\noindent
\emph{Propositon A} \newline \newline If a  nonrandom cipher has
nondegeneracy distance $n_d$, then it is broken by a
known-plaintext attack with data length $n=n_d$. $n=n_d$ is also
the smallest $n$ for which the cipher is broken with probability
1.
\newline \newline
\emph{Proof:} For any three joint random vectors $X_n$, $Y_n$,
$K$, we have the identity
\begin{equation} \label{A1}
H(Y_n|X_n)+H(K|X_n Y_n)=H(K|X_n)+H(Y_n|KX_n).
\end{equation}
For a nonrandom cipher, $H(Y_n|KX_n)=0$. In general, $H(K|X_n)
\leq H(K)$. Thus, $H(K|X_n Y_n)=0$ at any $n$ satisfying
Eq.~(\ref{broken}) or Eq.~(\ref{nondegdist}) and vice versa. From
its definition, $n_d$ is thus the smallest data length at which
the key is found for any given $x_n$.
\newline \newline
A similar result clearly holds for chosen plaintext attacks. Note
that if we consider the equation
\begin{equation} \label{A2}
H(X_n|Y_n)+H(K|X_n Y_n)=H(K|Y_n)+H(X_n|KY_n),
\end{equation}
then under (\ref{decryption}), a random cipher is broken by a
known-plaintext attack if $H(X_n|Y_n)$ satisfies the Shannon limit
(\ref{shannonlimit}) with equality. However, if one is satisfied
with using entropies as quantitative measures of security, one may
have a situation where
\begin{eqnarray} \label{A3}
\inf_n H(X_n|Y_n) = \lambda_1 H(K), \\ \label{A4} \inf_n
H(K|X_nY_n)= \lambda_2 H(K), \\ \label{A5} \inf_n H(K|Y_n)
=(\lambda_1 + \lambda_2) H(K),
\end{eqnarray}
under the constraint (\ref{A2}) where $0<\lambda_1,\lambda_2<1$
and $\lambda_1+\lambda_2 \leq 1$. (\ref{A3}-\ref{A5}) may still
provide satisfactory levels of security if $|K|$ is long enough,
and if Eve's information on the data bounded by $\lambda_1 H(K)$
does not help Eve to reduce her uncertainty on the rest of the
data below whatever designed level (We cannot enter into a
detailed discussion, as one problem of using entropies as
quantitative measures of security shows up here). While we have
not given any specific random cipher with such characteristics
proven, it has not been ruled out either. On the other hand, if
redundant key use or degeneracy is avoided for nonrandom ciphers,
then Proposition A applies. A detailed development will be given
elsewhere.
\end{document}